\documentclass{PoS}

\usepackage{graphicx}
\usepackage{amsmath}
\usepackage{amsfonts}
\usepackage{subfigure}
\usepackage{wrapfig}
\usepackage{epsfig}
\usepackage{afterpage,float}
\usepackage{cite}
\usepackage{ifpdf} 

\newcommand{\be}{\begin{equation}}
\newcommand{\ee}{\end{equation}}
\newcommand{\bea}{\begin{eqnarray}}
\newcommand{\eea}{\end{eqnarray}}

\def\half{{\textstyle{1\over2}}}

\def\bar{\overline}

\def\tilde{\widetilde}

\def\half{{\scriptstyle \raise.15ex\hbox{${1\over2}$}}}

\newcommand{\beq}{\begin{equation}}
\newcommand{\eeq}{\end{equation}}

\newcommand{\real}{\relax{\rm I\kern-.18em R}}

\title{The sextet gauge model, light Higgs, and the dilaton}

\ShortTitle{The sextet gauge model, light Higgs, and the dilaton}

\author{Zolt\'an Fodor\\
        Department of Physics, University of Wuppertal\\
        Gau$\beta$strasse 20, D-42119, Germany\\
        J\"ulich Supercomputing Center, Forschungszentrum
        J\"ulich, D-52425 J\"ulich, Germany\\
        Email: \email{fodor@bodri.elte.hu}}

\author{Kieran Holland\\
        Albert Einstein Center for Fundamental Physics, Institute for
        Theoretical Physics, \\
        Bern University, Sidlerstrasse 5, CH-3012 Bern, Switzerland\\
       Department of Physics, University of the Pacific,
        3601 Pacific Ave, Stockton CA 95211, USA\\
        Email: \email{kholland@pacific.edu}}

\author{\speaker{Julius Kuti }\\
        Department of Physics 0319, University of California, San Diego\\
        9500 Gilman Drive, La Jolla, CA 92093, USA\\
        E-mail: \email{jkuti@ucsd.edu}}

\author{D\'aniel N\'ogr\'adi\\
        Institute for Theoretical Physics, E\"otv\"os University\\
        H-1117 Budapest, Hungary\\
        Email: \email{nogradi@bodri.elte.hu}}

\author{Chris Schroeder\\
        Physical Sciences Directorate, Lawrence Livermore National Laboratory\\
        Livermore, California 94550, USA\\
        E-mail: \email{schroeder10@llnl.gov} }
        
\author{Chik Him Wong\\
        Department of Physics 0319, University of California, San Diego\\
        9500 Gilman Drive, La Jolla, CA 92093, USA\\
        E-mail: \email{rickywong@physics.ucsd.edu} }

\abstract{
The frequently discussed strongly interacting
gauge theory with a fermion flavor doublet in the two-index symmetric (sextet) representation
of the SU(3) color gauge group is investigated~\cite{Fodor:2012ty}. 
The chiral condensate and the mass spectrum 
are consistent with chiral symmetry breaking $(\chi{\rm SB})$ at vanishing fermion mass.
In contrast, sextet fermion mass deformations of 
spectral properties are not consistent with leading conformal scaling behavior near the critical surface 
of a conformal theory.
A recent paper~\cite{DeGrand:2012yq} which could not resolve
the conformal fixed point of the gauge coupling from the slowly walking scenario 
of a very small nearly vanishing $\beta$-function is not in  conflict
with $\chi{\rm SB}$ reported here. 
A light Higgs impostor could emerge as the 
dilaton from spontaneous symmetry breaking of scale invariance
or, without the dilaton mechanism, as a composite state. 
}

\FullConference{The 30 International Symposium on Lattice Field Theory - Lattice 2012,
		June 24-29, 2012\\
		Cairns, Australia}

\begin{document}


\section{Introduction}

The new Higgs-like particle with decay modes 
not far from that of
the Standard Model brings new focus and clarity to the search for theoretical
frameworks.
One example is the light dilaton as a pseudo-Goldstone particle of spontaneous  breaking of 
scale invariance that has been featured in recent phenomenological discussions as a viable
interpretation of the discovery.
Nearly conformal gauge theories serve as  theoretical laboratories for credible realizations
of this scenario (this short report is based on~\cite{Fodor:2012ty} which provides
references to the history of the field including recent work).
We investigate here
a candidate theory with a fermion flavor doublet in the two-index symmetric (sextet) representation
of the SU(3) color gauge group close to the conformal window with a small beta function, 
as suggested by results of a recent paper~\cite{DeGrand:2012yq}. 
With the limited statistical accuracy of the nearly vanishing $\beta$-function 
the existence of a conformal fixed point gauge coupling remains unresolved from the alternative slowly walking scenario.
Consistency with $\chi{\rm SB}$, reported here for small fermion mass deformations,
would require the sextet model 
to remain just below the conformal window with a very small non-vanishing $\beta$-function (see, also~\cite{Kogut:2010cz,Kogut:2011ty}). 
In this case the model which exhibits the simplest composite Higgs mechanism leaves open the possibility
of a light scalar state  emerging as the 
pseudo-Goldstone dilaton state from spontaneous symmetry breaking of scale invariance. 
Even if scale symmetry breaking is entangled with
$\chi{\rm SB}$  without dilaton interpretation, 
a light Higgs-like composite state can emerge  close to the conformal window.
We outline a  new lattice Higgs project to resolve these important problems.

\section{Computational strategy and lattice simulations}

Probing $\chi{\rm SB}$, and conformal behavior for comparison, we extrapolate the spectrum 
to infinite volume at fixed fermion mass $m$. In large volumes the leading 
finite size corrections  are exponentially small and dominated by the lowest state of the spectrum which
is expected to have pion quantum numbers.
From the mass spectrum, extrapolated to  infinite volume, we can probe the pattern of $\chi{\rm SB}$
when small fermion mass deformations are simulated close to the massless limit. We also probe the hypothesis of
mass deformed conformal scaling behavior and find results strongly favoring the $\chi{\rm SB}$ hypothesis.

A new analysis is presented at gauge coupling $\beta=3.2$ 
based on a subset of runs in the fermion mass range ${\rm m=0.003-0.010}$ on
$24^3\times48$,  $28^3\times56$,  and $32^3\times64$ lattices. Five fermion masses
at  m=0.003, 0.004, 0.005, 0.006, 0.008 are used in most fits probing the two hypotheses.
A very large and expensive $48^3\times96$ 
run was added recently at ${\rm m=0.003}$ to control finite size effects.
Simulation results at $\beta=3.25$ were also obtained in the mass range ${\rm m=0.004-0.008}$ on
$24^3\times48$,  $28^3\times56$, and $32^3\times64$ lattices. 
We have used the tree-level Symanzik-improved gauge action with stout smeared staggered fermions 
for all results reported here.
The normalization of the lattice coupling $\beta$ and details of the simulations are discussed in~\cite{Fodor:2012ty}.

To control finite size effects, infinite-volume extrapolations were performed for the lowest state in the spectrum 
\begin{figure}[h!]
\begin{center}
\begin{tabular}{ccc}
\includegraphics[height=3.8cm]{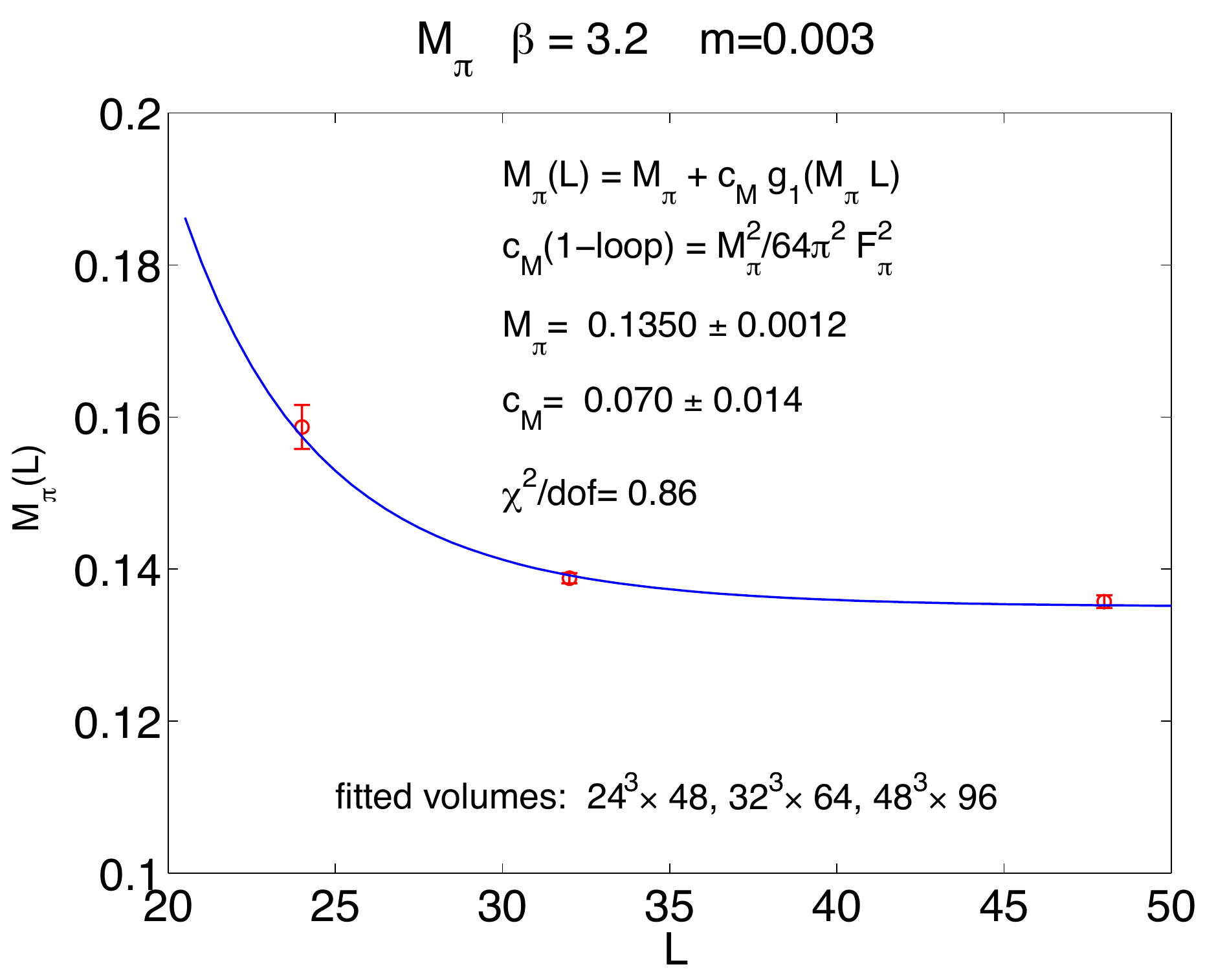}&
\includegraphics[height=3.8cm]{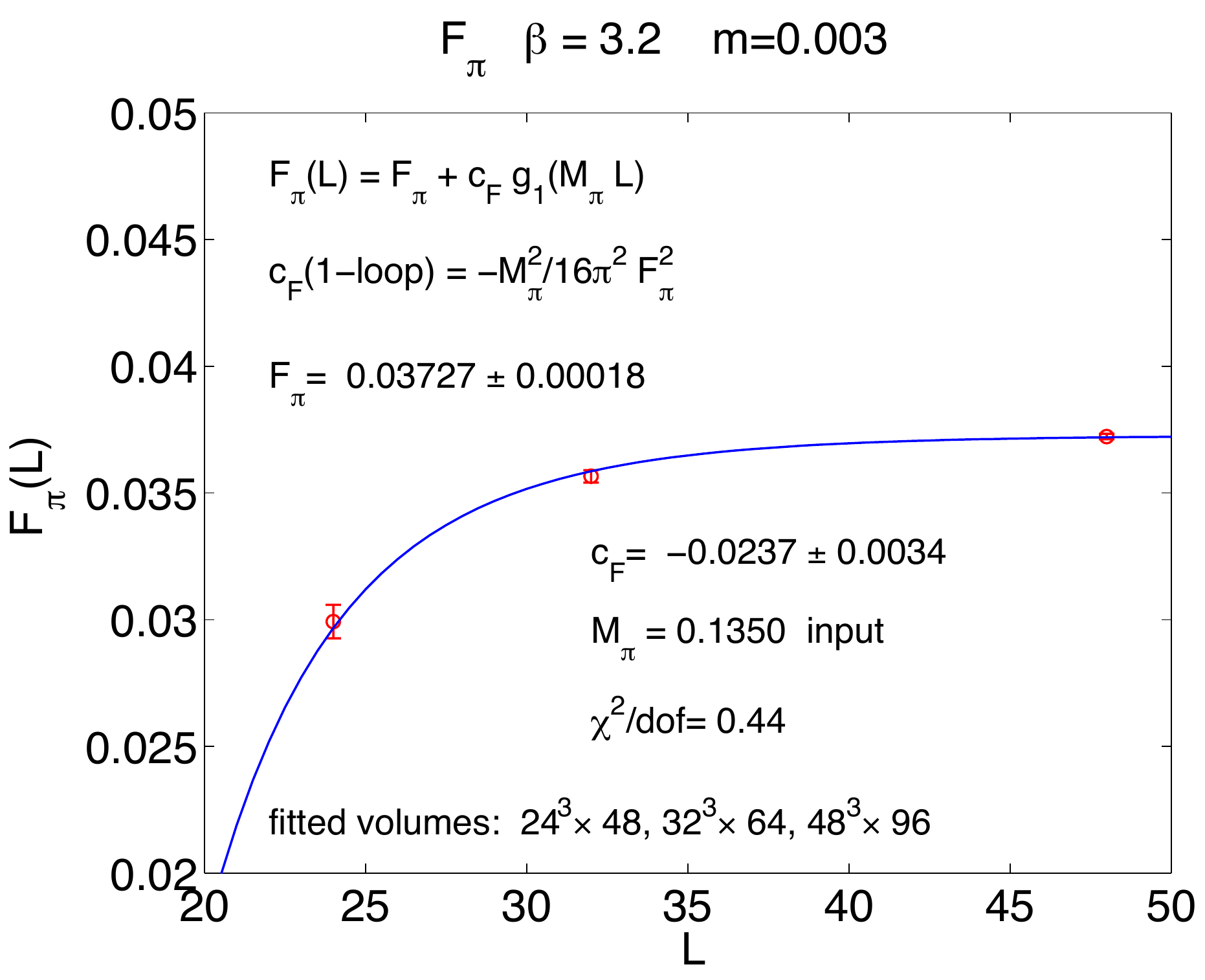}&
\includegraphics[height=3.8cm]{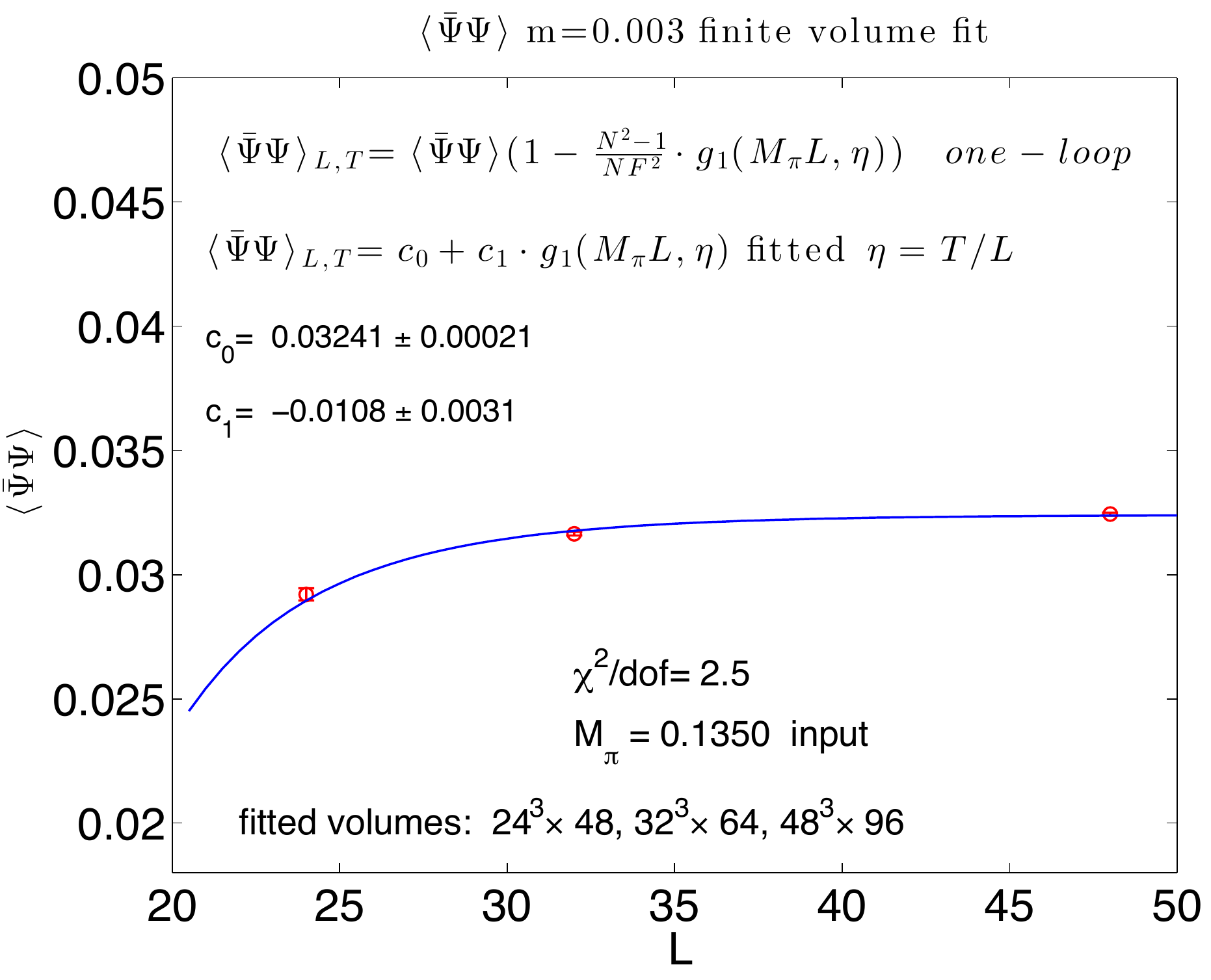}
\end{tabular}
\end{center}
\vskip -0.2in
\caption{\footnotesize  Finite volume dependence at the lowest fermion mass for $\beta=3.2$.
The form of $\tilde g_1(\lambda,\eta)$ is a complicated infinite sum which contains Bessel
functions and requires numerical evaluation~\cite{Gasser:1986vb}. Since we are not in the chiral log regime, the prefactor of
the $\tilde g_1(\lambda,\eta)$ function was replaced by a fitted coefficient. The leading term of  the function
$\tilde g_1(\lambda,\eta)$ is a special exponential Bessel function $K_1(\lambda)$ which dominates in the simulation range.}
\vskip -0.1in
\label{fig:sextetInfVol}
\end{figure}
with pion quantum numbers, the related decay constant $F_\pi$,
and the chiral condensate $\langle\bar\psi\psi\rangle$. They
are shown in Figure~\ref{fig:sextetInfVol} where $\tilde g_1(\lambda,\eta)$ describes finite volume corrections 
from the exchange of the lightest pion state with  $\lambda=M_\pi L$ and lattice aspect ratio $\eta=T/L$,
similarly to what was introduced in~\cite{Leutwyler:1987ak}.  
The fitting procedure approximates the leading treatment of  the pion which wraps around the finite volume,
whether in chiral perturbation theory ($\chi{\rm PT}$), or in L\"uscher's non-perturbative finite size analysis~\cite{Luscher:1985dn}. 
This equivalence relaxes the requirement on the fitted parameters $c_M$,$c_F$,$c_1$ to agree with 1-loop  
$\chi{\rm PT}$ as long as the pion is the lightest state dominating the finite volume corrections.
The infinite-volume limits of $M_\pi$, $F_\pi$,  and $\langle\bar\psi\psi\rangle$ for  $m=0.003$ at $\beta=3.2$
were determined self-consistently from the fitting procedure. Similar fits were applied to other composite states.
The value of $M_\pi$ in the fit of the left plot in Figure~\ref{fig:sextetInfVol} was 
determined from the highly non-linear fitting function and used as input in the other two fits.
Based on the fits at $m=0.003$,  
the results are within one percent of the infinite-volume  limit at $M_\pi L= 5$.
In the fermion mass range $m \geq 0.004$ the condition $M_\pi L> 5$ is reached at $L=32$
and we will interpret the results from the  $32^3\times64$ runs for $m \geq 0.004$
as infinite-volume  behavior in mass deformed chiral and conformal analyses.

\section{The chiral condensate} 
Our simulations show that the chiral condensate $\langle \bar{\psi}\psi\rangle$ is consistent with $\chi{\rm SB}$ 
and remains non-vanishing in the massless fermion limit.
The cutoff dependent UV contributions are identified at finite fermion mass.
The linear mass term $c_1(a)\cdot m$ is a quadratically divergent UV contribution 
$\approx a^{-2}\cdot m$ with lattice cutoff $a$.
There is also a very small
third-order UV term  $c_3(a)\cdot m^3$ without power divergences which is hard to detect for small $m$ and has 
not been tested within the accuracy of the simulations. 
IR finite contributions are identified in the chiral expansion of the condensate. There is an $m$-independent constant
term which is proportional to $ B F^2$, a linear term proportional to $ B^2\cdot m$, a quadratic term $\sim B^3F^{-2}\cdot m^2$, and higher 
order terms, in addition to
logarithmic corrections generated from chiral loops. 
The expansion in the fermion mass is expressed in terms of low energy constants 
of chiral perturbation theory, like $B$ and $F$~\cite{Bijnens:2009qm}. 
A more complete description of our fitting functions is given in~\cite{Fodor:2012ty}.
\begin{figure}[h!]
\begin{center}
\begin{tabular}{cc}
\includegraphics[width=5cm]{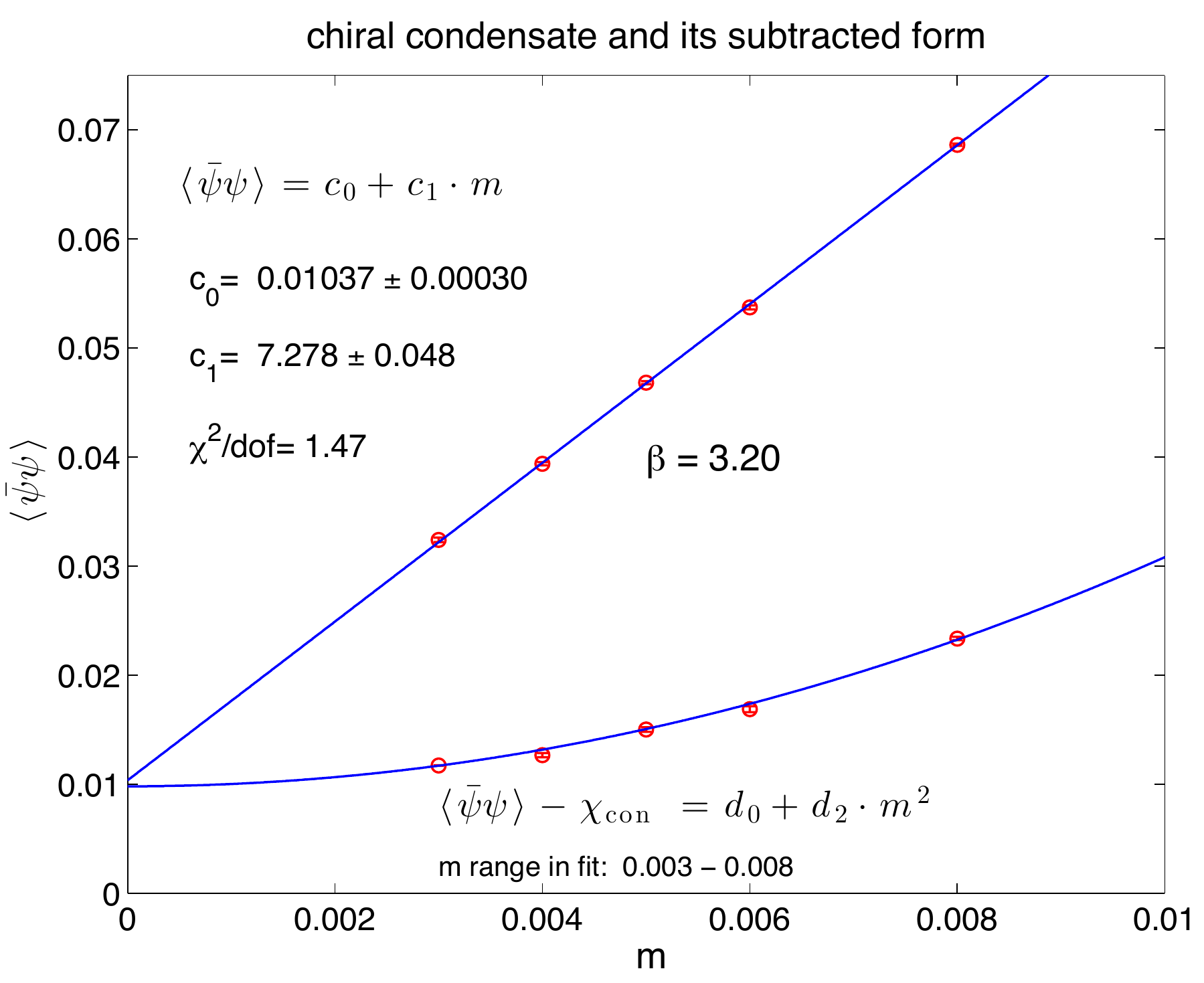}&
\includegraphics[width=5cm]{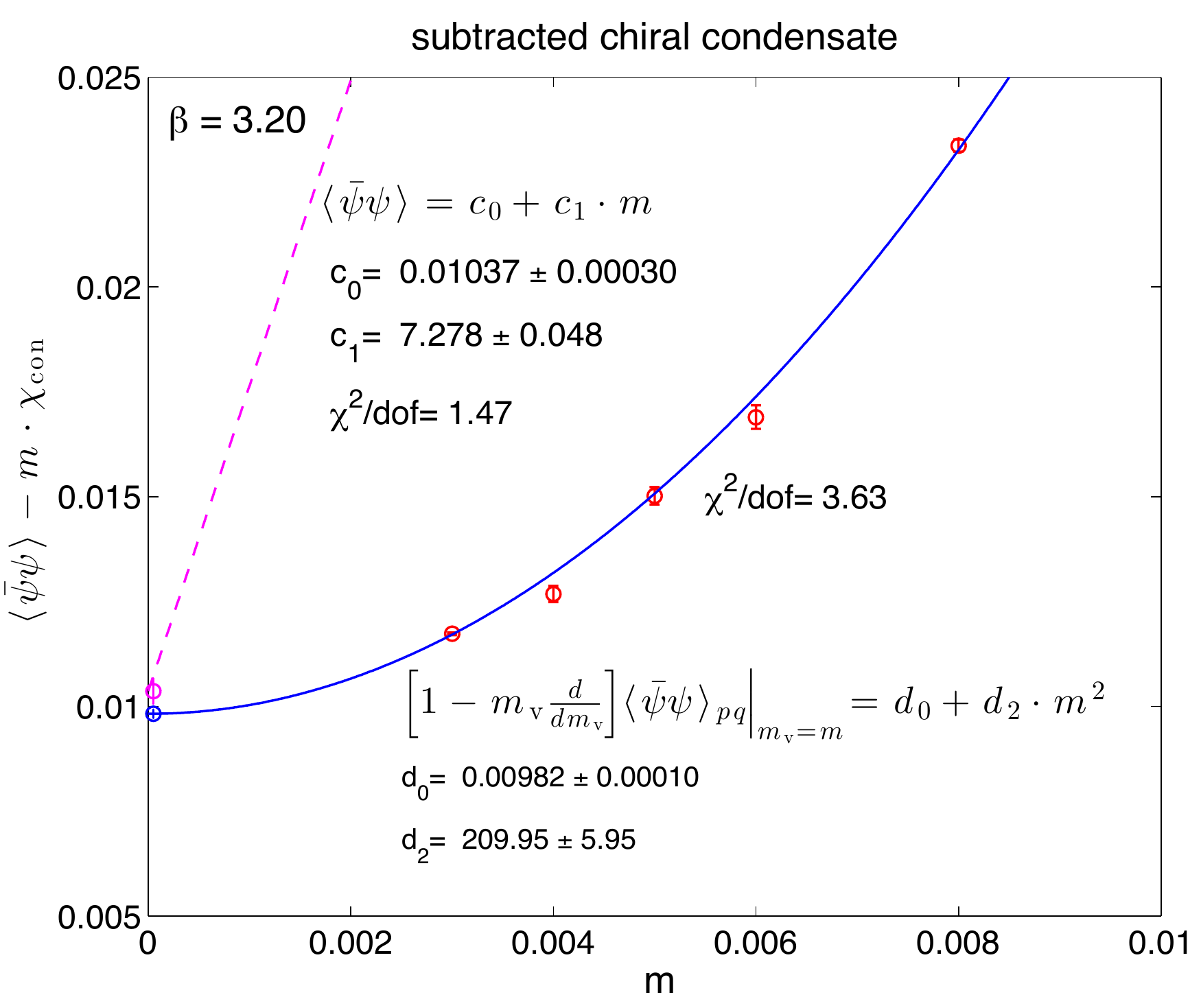}
\end{tabular}
\end{center}
\vskip -0.2in
\caption{\footnotesize The chiral condensate and its reduced form with subtracted derivative  (both have to converge  
to the same chiral limit) are shown in the left plot with linear fit to the condensate. The data without derivative subtraction cannot
detect higher order fermion mass terms with significant accuracy. 
 The fit to the reduced form
with subtracted derivative is defined in~\cite{Fodor:2012ty} and shown in the magnified right plot. 
A linear term is not included in this fit since the subtracted derivative form approximately eliminates it. 
The value of  $d_0$ at $m=0$  is shown to be consistent
with the direct determination of $c_0$ from the chiral limit of $\langle \bar{\psi}\psi\rangle$.
The consistency is very reassuring since the two results are derived from independent determinations.
For $m=0.003$ the data from infinite-volume extrapolation were used in the fit. 
As we explained earlier, at higher $m$ values the largest volume 
$32^3\times 64$ runs were used for the condensate and its derivative subtraction.}
\label{fig:PbPsextet}
\end{figure}
We used two independent methods for the determination of the chiral condensate in the
massless fermion limit. In the first method fits were made directly to 
 $\langle \bar{\psi}\psi\rangle$ with constant and linear terms in the fitted function.
Quadratic 
and third order  terms are hard to detect within the accuracy of the data. 
The result is shown in the left plot of Figure~\ref{fig:PbPsextet}.  
When the quadratic term is added to the fit,
the massless intercept  $c_0=\langle \bar{\psi}\psi\rangle_{m=0}$ from the quadratic fit agrees with the one from the linear fit 
and the quadratic fit coefficient in $c_2\cdot m^2$
is zero within fitting error. 
For an independent determination, we also studied the subtracted chiral condensate operator defined with the help
of the connected part $\chi_{conn}$ of the chiral susceptibility $\chi$.
Once the derivative term is subtracted, the first non-perturbative IR contribution, 
quadratic in $m$, is better exposed. 
The two independent determinations give consistent non-vanishing fit results in the massless chiral limit
as shown in Figure~\ref{fig:PbPsextet}.

\section{Spectral tests of the $\chi {\rm SB}$ hypotheses}


The chiral Lagrangian describes the low energy theory of
Goldstone pions and non-Goldstone pions in the staggered lattice fermion formulation.
It can be used as an effective tool probing the $\chi{\rm SB}$  hypothesis at finite fermion masses
including extrapolation to the massless chiral limit.
\begin{figure}[h!]
\begin{center}
\begin{tabular}{cc}
\includegraphics[height=4cm]{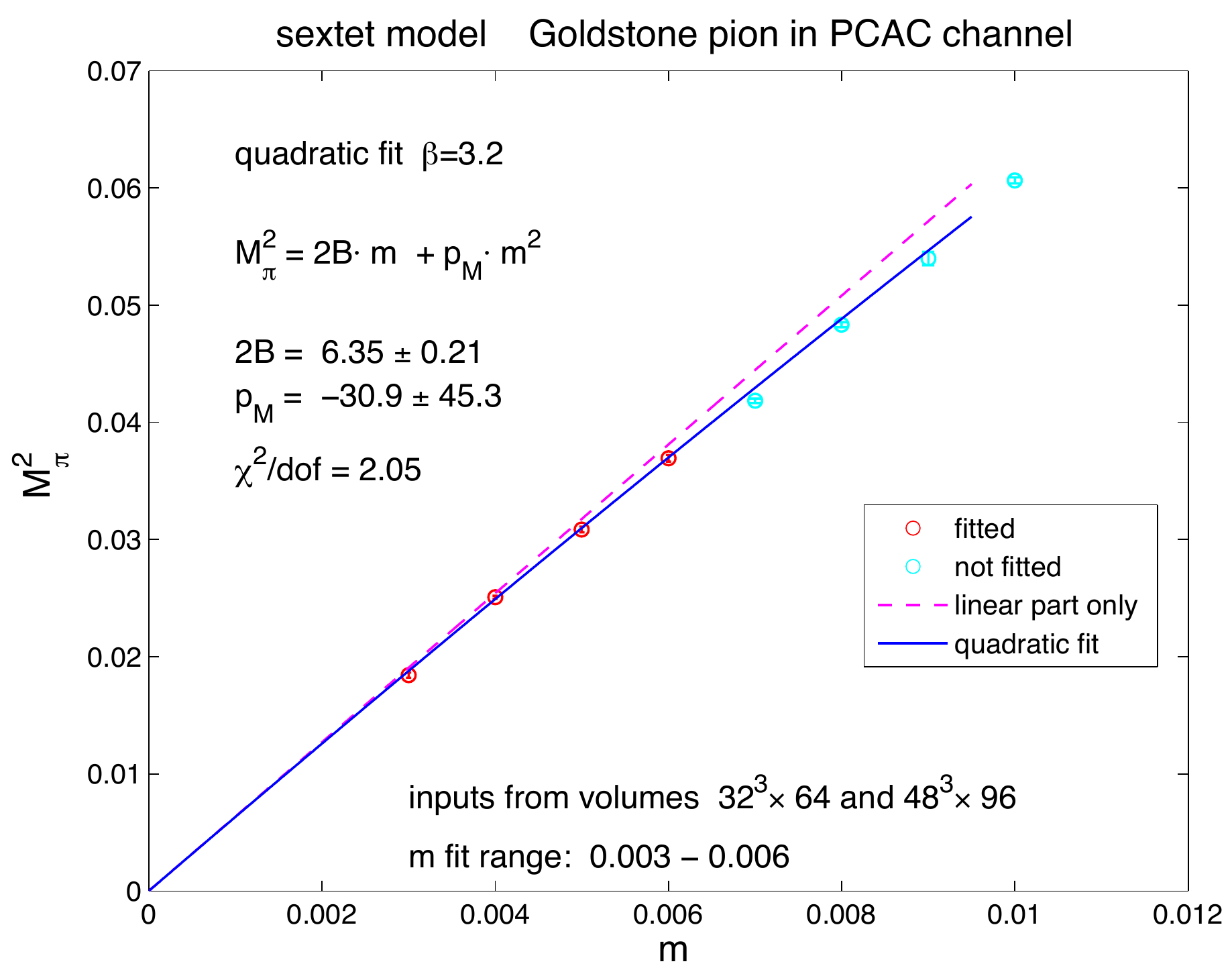}&
\includegraphics[height=4cm]{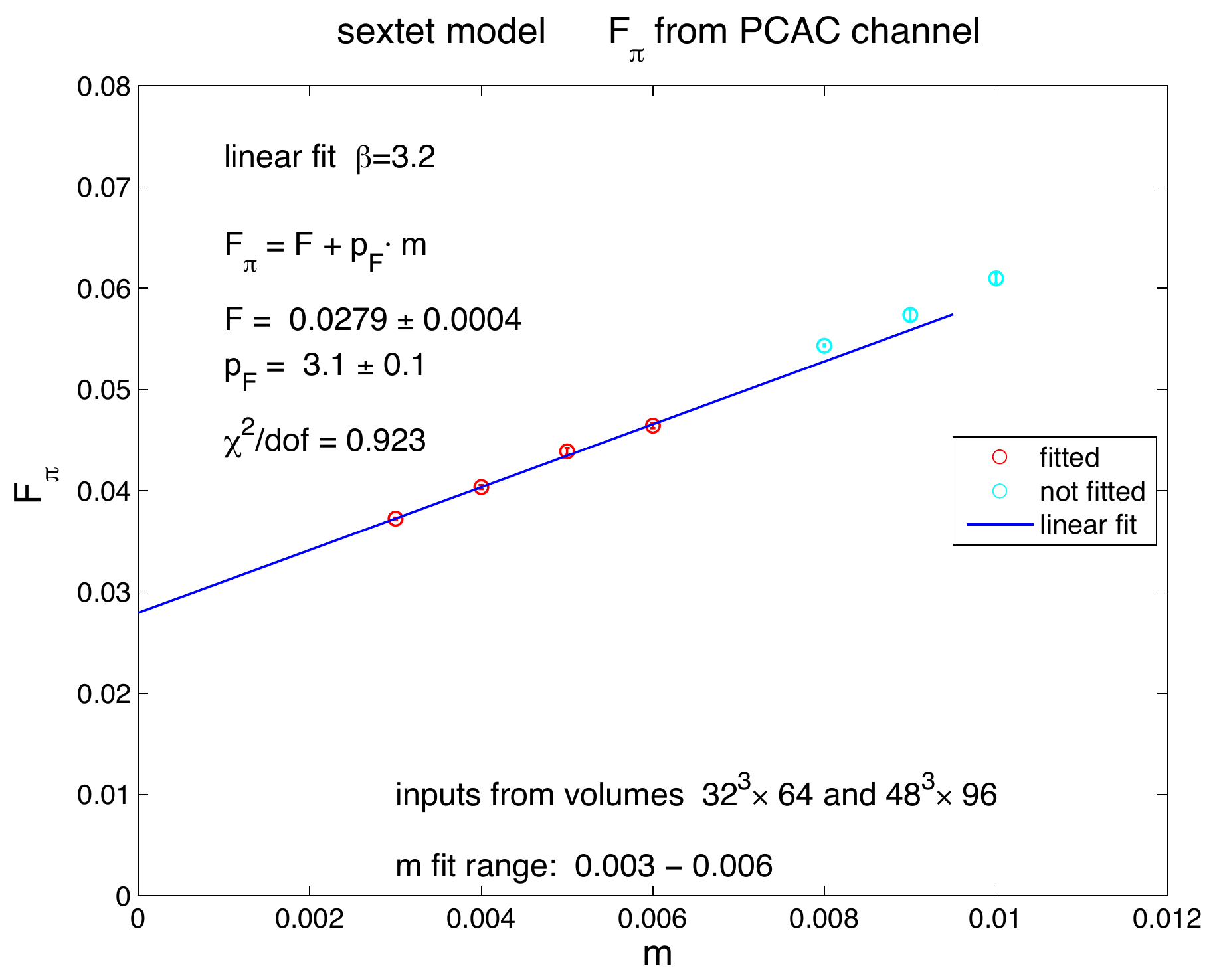}
\end{tabular}
\end{center}
\vskip -0.2in
\caption{\footnotesize  Polynomial fits from the analytic mass dependence of the chiral Lagrangian without logarithmic 
loop corrections are shown for the Goldstone pion and  $F_\pi$. The dashed line in the left plot for the Goldstone pion
shows the leading linear contribution. The data point at $m=0.003$ is determined from the $48^3\times 96$ lattice and the other
fitted points are based on the $32^3\times 64$ runs.}
\label{fig:MpiFpi}
\end{figure}
Close to the chiral limit, the  pion spectrum and the pion decay constant $F_\pi$ 
are organized in  powers of the fermion mass $m$ which is an input parameter in the simulations.
Chiral  log corrections to the polynomial terms are
generated from pion loops~\cite{Gasser:1983yg}. Their analysis will
require an extended dataset with high statistics. 

In Section 2 we presented results of infinite-volume extrapolations.
Based on these observations, in fits to the observed pion spectrum and $F_\pi$, 
we will use infinite-volume extrapolation at $m=0.003$ and treat the $32^3\times64$ runs for $m \geq 0.004$
as if the volume were infinite.
In Figure~\ref{fig:MpiFpi} we used the local pion correlator with noisy sources to extract $M_\pi$ and $F_\pi$.
The correlator is tagged as the PCAC channel since the PCAC relation, based on axial Ward identities, holds for this correlator 
and the decay constant $F_\pi$ can
be directly determined from the residue of the pion pole.
Based on the analytic fermion mass dependence of the chiral Lagrangian, and using the lowest four fermion masses,
good polynomial fits were obtained for $M_\pi$ and $F_\pi$ as shown in Figure~\ref{fig:MpiFpi} with
fitting functions $M^2_{\pi}  = 2B\cdot  m + p_M \cdot  m^2$ and 
$F_{\pi } = F + p_{F}\cdot m$ .
The parameters $B$ and $F$ are defined in the two leading terms of the chiral Lagrangian~\cite{Gasser:1983yg}. 
In this simple fitting procedure $B$ is set from the fit to $M^2_\pi$ and $F$ is 
set independently from the fit to $F_\pi$.
The fit parameters $p_M$ and $p_F$ describe the respective leading power corrections in $m$
to the pion mass and decay constant. 
Limited to a single lattice spacing only, the accuracy of our dataset is not sufficient for the robust determination of chiral log
corrections that will provide important consistency check for
$B$ and $F$ in our future analysis of an extended dataset. 
This analysis requires rooted and partially quenched staggered perturbation theory
at finite lattice spacing for simultaneous fits of $M_\pi$ and $F_\pi$ 
with a consistent pair of cutoff-dependent $F$ and $B$ values~\cite{Aubin:2003mg}.
%
%
We made the first step in this direction by adding a new run set
to our database at $\beta=3.25$. 
Reduction in taste breaking is significant at  $\beta=3.25$ with smaller lattice spacing. Our
staggered perturbation theory analysis will be presented elsewhere.

\section{Spectral tests of the conformal scaling hypothesis}
Under the conformal scaling hypothesis, the mass $M_\pi$
and the decay constant $F_\pi$ are given at leading order by $M_\pi = c_M\cdot m^{1/1+\gamma}$ and $F_\pi = c_F\cdot m^{1/1+\gamma}$.  
The coefficients $c_M$ and $c_F$
are channel specific but the exponent $\gamma$ is  universal in all channels~\cite{DelDebbio:2010ze}. 
The leading scaling form sets in for small $m$ values,
close to the critical surface. According to the hypothesis, there is an infrared conformal fixed point on the critical surface which controls
the conformal scaling properties of small mass deformations. 
All masses of the spectrum can be subjected to similar conformal
scaling tests, but we will mostly focus on accurate data in the  $M_\pi$ and $F_\pi$ channels. 

When $M_\pi$ and $F_\pi$ are fitted {\em separately} in the range of 
 the four lowest fermion masses  closest to the critical surface, we get reasonable $\chi^2$ values for the fits. 
 However, the incompatibility of the fitted $\gamma$ values
disfavors the hypothesis,  inconsistent
with mass deformed conformal behavior.
\begin{figure}[h!]
\begin{center}
\begin{tabular}{cccc}
\includegraphics[height=2.8cm]{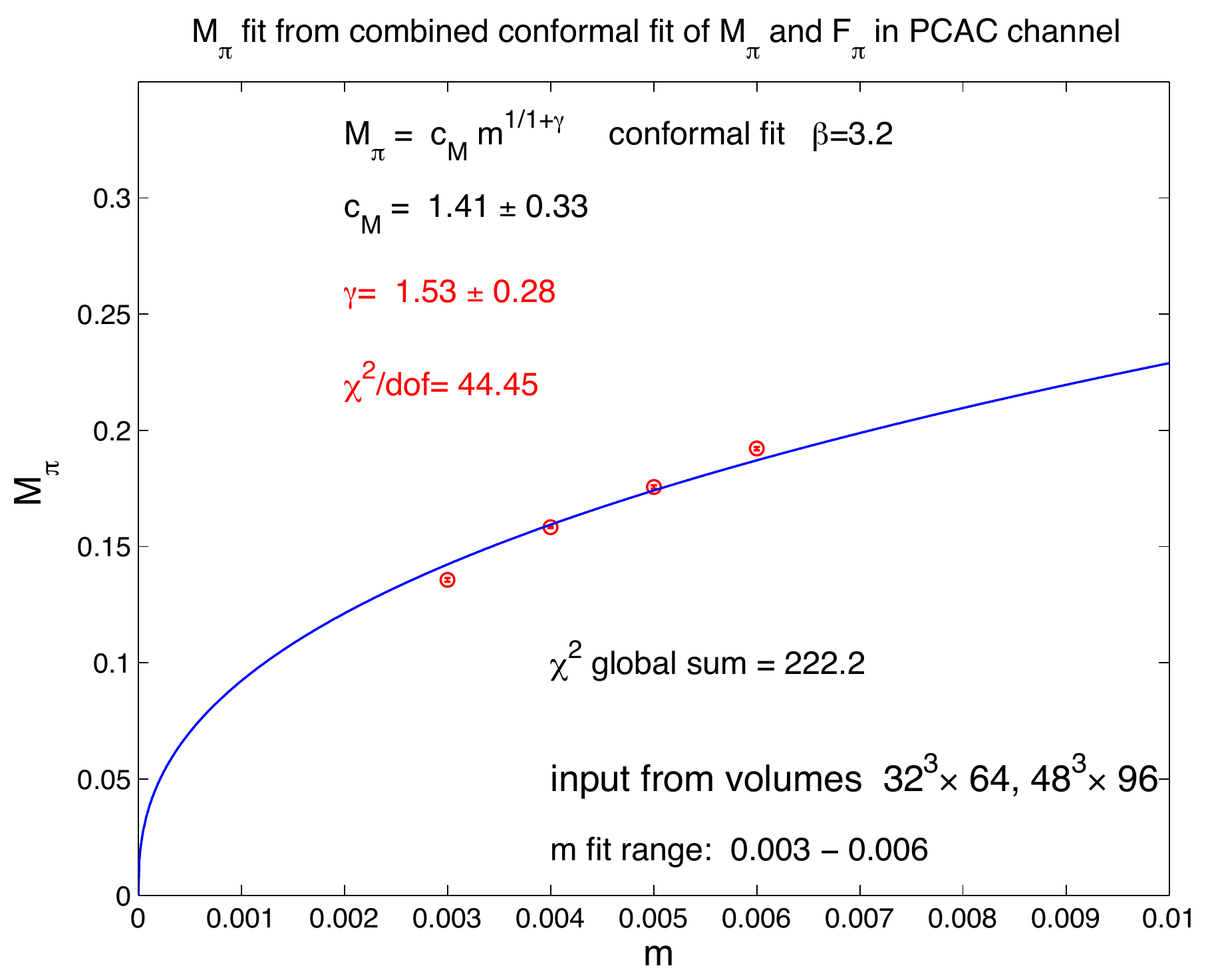}&
\includegraphics[height=2.8cm]{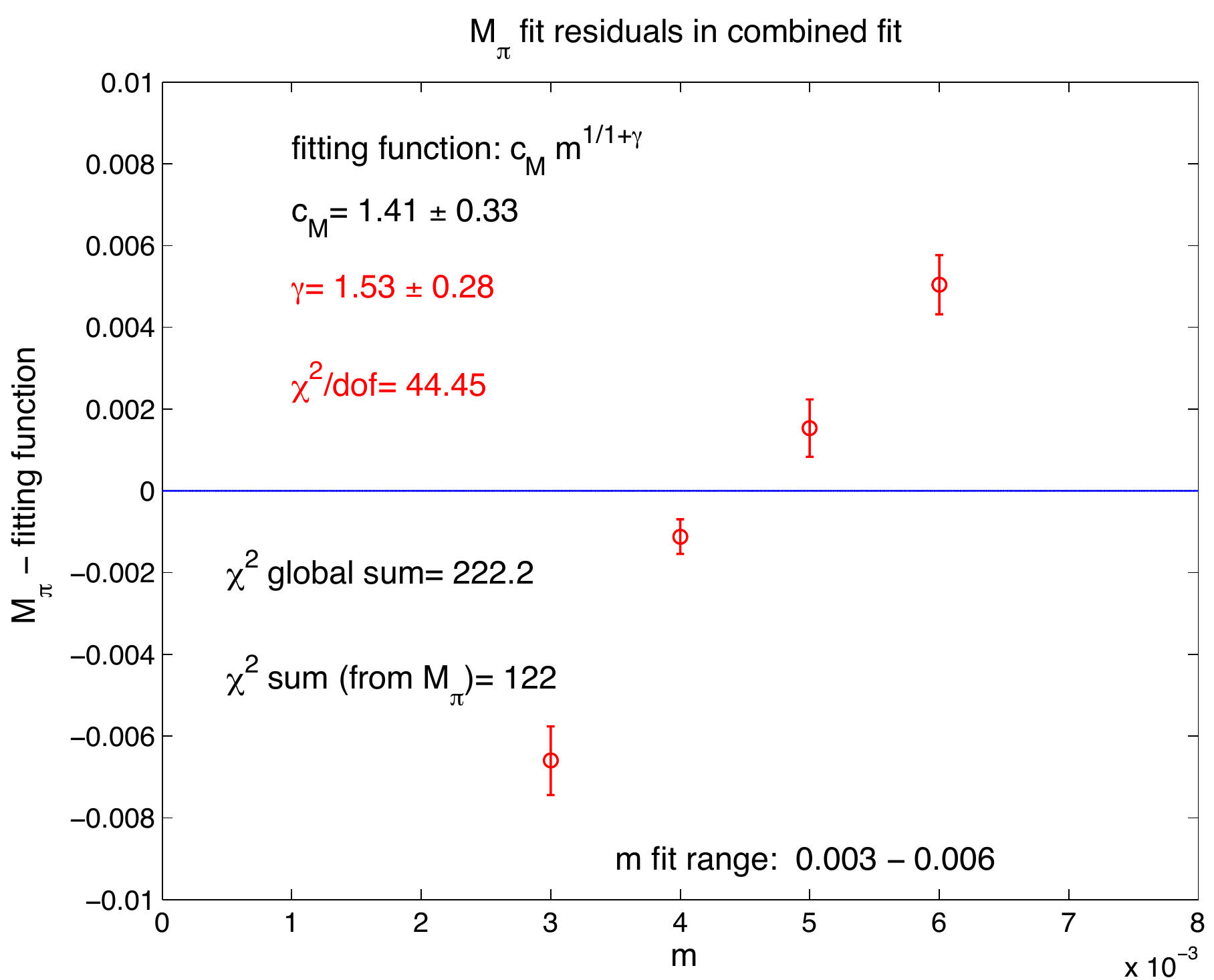}&
\includegraphics[height=2.8cm]{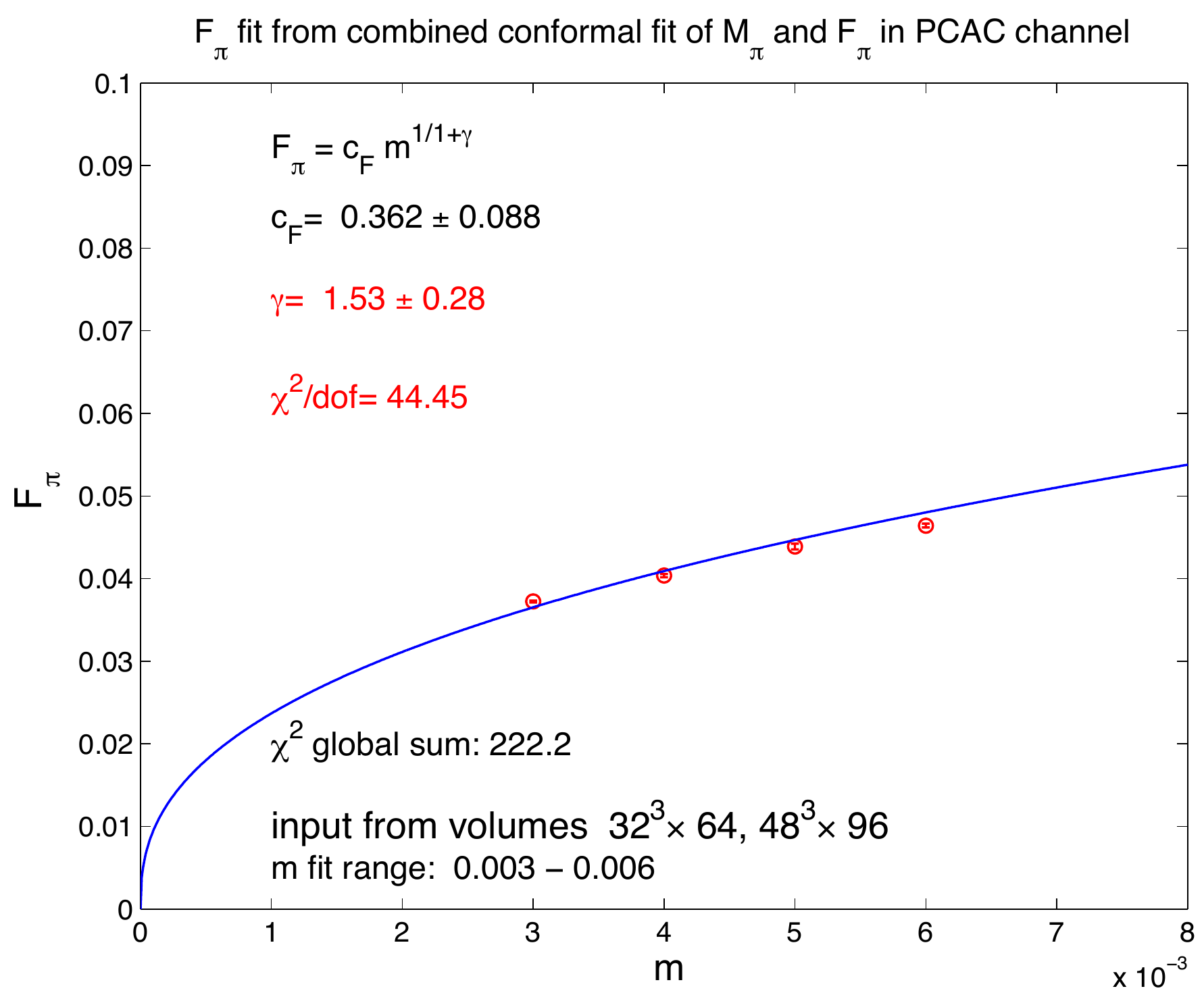}&
\includegraphics[height=2.8cm]{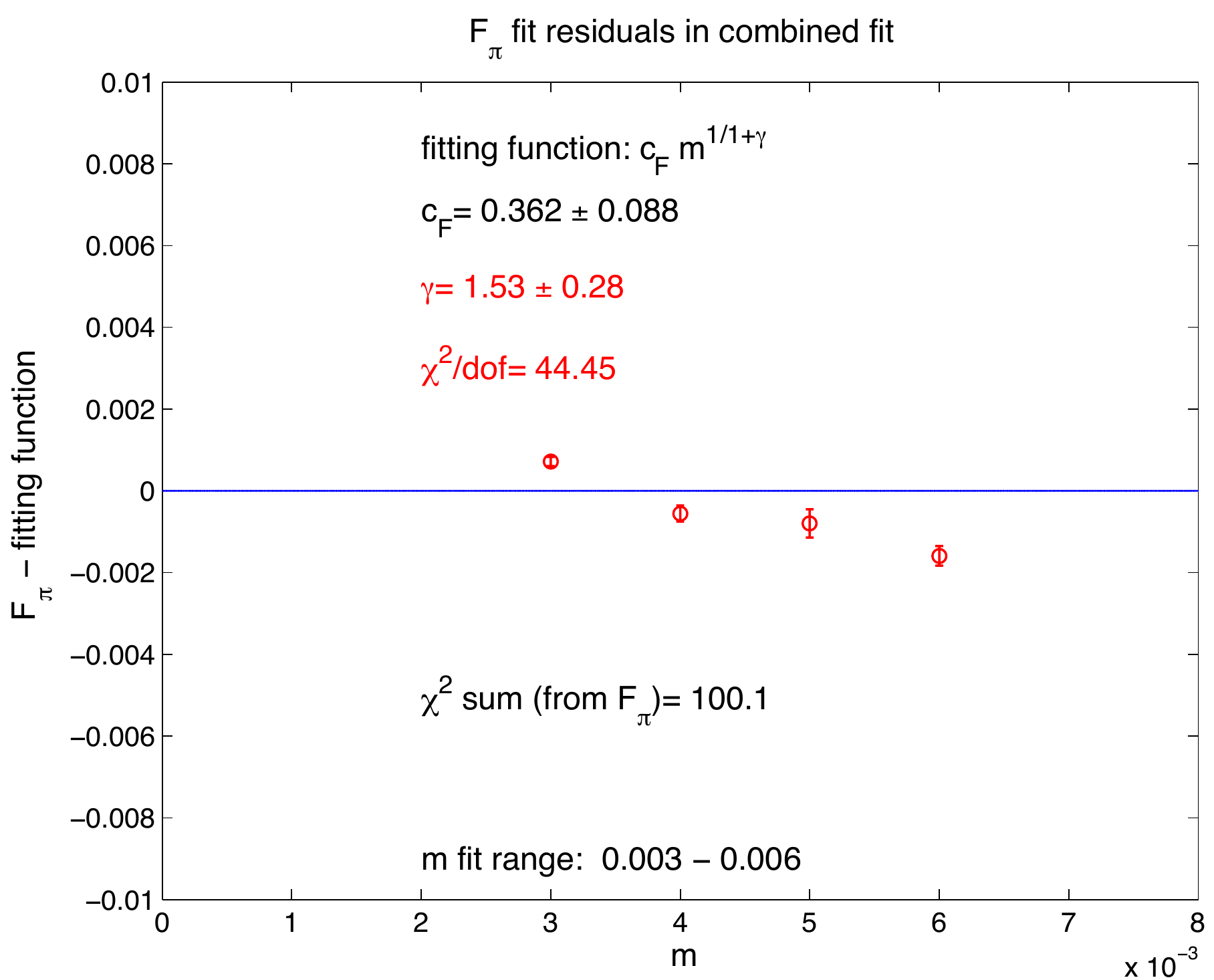}
\end{tabular}
\end{center}
\vskip -0.1in
\caption{\footnotesize The first plot shows the simultaneous conformal fit result for
the pion mass, while the second displays the $M_\pi$ residuals. The last two
plots show the simultaneous fit result for the pion decay constant and the
$F_{\pi}$ residuals.
The combined fit forces $\gamma=1.53(28)$
with an unacceptable ${\rm \chi^2/dof}$ of 44.5.
}
\label{fig:sextetConformTest2}
\vskip -0.1in
\end{figure}
 The conflicting simultaneous fits to  universal conformal form with the same $\gamma$  
 for the Goldstone pion and the $F_\pi$ decay constant
 are illustrated  in Figure~\ref{fig:sextetConformTest2}. Fitting to the pion mass separately requires $\gamma=1.040(73)$ while the 
separate $F_\pi$ fit is forcing  $\gamma=2.20(15)$. In the combined fit they compromise with  $\gamma=1.53(28)$
and the unacceptable ${\rm \chi^2/dof}$ of 44.5.
It is important to note that the exponent $\gamma$  for the fit to $M_\pi$ only is what 
$\chi{\rm SB}$ would prefer. The separate conformal exponent $\gamma$  for $F_\pi$ is large to force 
to the origin the linear string of data which extrapolate to a finite constant in $\chi{\rm SB}$.
This creates conflict with the universal exponent $\gamma$  in the conformal analysis.   

From the tests we were able to perform, the sextet model is consistent with $\chi {\rm SB}$ in leading approximation
and inconsistent with conformal symmetry. 
It will require further investigations to show that subleading effects cannot alter this conclusion. It is particularly 
important to strengthen the analysis of the $\chi{\rm SB}$ hypothesis reaching the necessary accuracy to identify 
chiral log effects.
We will also consider comprehensive conformal finite size scaling (FSS) tests which do not rely on 
infinite-volume extrapolation in the scaling fits.  
It remains difficult to reconcile  $\chi{\rm SB}$ and large exponents in the fermion mass dependence with the low value of $\gamma$ 
defined by the chiral condensate 
using the Sch\"odinger functional for massless fermions~\cite{DeGrand:2012yq}.

\section{The new sextet Higgs project}

\begin{figure}[h!]
\begin{center}
\begin{tabular}{c}
\includegraphics[height=5cm]{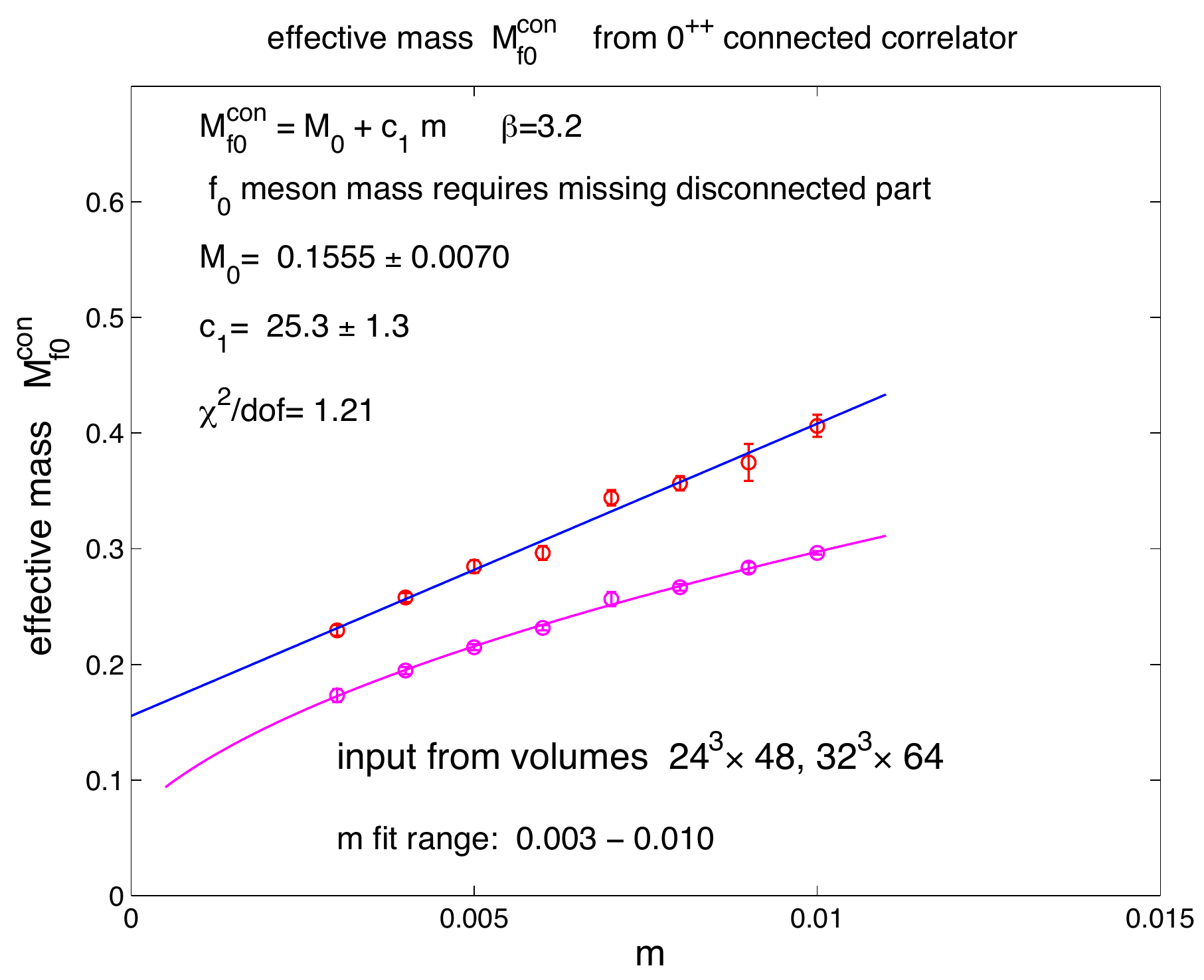}
\end{tabular}
\end{center}
\vskip -0.2in
\caption{\footnotesize  The linear fit is shown to the mass of the scalar $0^{++}(f_0)$  meson from the 
connected part of correlator I in Table 1 of~\cite{Ishizuka:1993mt}. For comparison, the scPion which is the opposite parity
state in the staggered correlator is also plotted (at $m=0$ the mass of the non-Goldstone scPion will vanish in the continuum). 
Gluon annihilation and mixing of the $f_0$ state with a low mass $0^{++}$
glueball state could generate a light scalar mass close to the conformal window.
The disconnected part of the staggered correlator and its mixing with the glueball correlator are required to resolve this issue.}
\vskip -0.1in
\label{fig:f0}
\end{figure}

Figure~\ref{fig:f0} shows the fermion mass dependence of the scalar $0^{++}(f_0)$ meson without including the disconnected
part of correlator I in Table 1 of~\cite{Ishizuka:1993mt}. The non-Goldstone scPion and $f_0$ are 
opposite parity states in this staggered correlator.
The quantum numbers of the $f_0$ meson match those of the $0^{++}$ state in the staggered correlator.
Close to the conformal window the  scalar  $0^{++}$ meson is not expected to be similar to the $\sigma$ particle of QCD.
If it turns out to be light, the full scalar state, including the disconnected gluon annihilation diagram and mixing with the $0^{++}$ 
glueball state, could replace the role of the 
elementary Higgs and act as the Higgs impostor. 
It is  very difficult to do the full calculation including the disconnected gluon annihilation diagram and mixing 
with the $0^{++}$ glueball state. This is the main part of our next generation sextet Higgs project. 

%

The linear  fit from the connected diagram is shown in Figure~\ref{fig:f0}. It has a non-zero intercept in the chiral limit with a mass more 
than five times $F$ so it corresponds to a heavy state and not a light Higgs candidate. 
Close to the conformal window it is reasonable to expect that the disconnected
diagram, after mixing with the $0^{++}$ glueball state will lead to a light scalar mass in $F$ units when the chiral limit is taken. 
The light scalar state after mixing remains a viable Higgs candidate.
Only new simulations can resolve the issue and perhaps eliminate this attractive scenario.

To study flavor-singlet mesons, we need to consider fermion loops which are disconnected (often called hairpin diagrams).
Flavor-singlet correlators have 
fermion-line connected and fermion-line disconnected contributions from the hairpin diagrams.
To evaluate disconnected quark loops with zero momentum, we need to sum over propagators from sources 
at each spatial location for a given time slice. 
To avoid the very costly $\mathcal{O}(V)$
inversions to compute  all-to-all propagators in lattice terminology,  random sources have to be used with  noise
reduction.

A  very interesting remaining challenge and complication is the existence of two types of distinct $0^{++}$ scalar states. 
One of them is the composite meson state and the other is the scalar glueball with the same $0^{++}$  quantum numbers.  
In dynamical sextet simulations, these two types of state will mix producing two scalar states. To resolve this will
require a well-chosen variational operator set with room left for a light
scalar state to emerge in the spectrum. It is also entirely possible that careful lattice calculations will shut down the light Higgs interpretation.

\section*{Acknowledgments}
\vskip -0.1in
We acknowledge support by the DOE under grant DE-FG02-90ER40546, by the NSF under grants 0704171 and 0970137, 
by the EU Framework Programme 7 grant (FP7/2007-2013)/ERC No 208740, and 
by the Deutsche Forschungsgemeinschaft grant SFB-TR 55. Computational resources were 
provided by USQCD at Fermilab and JLab, by the NSF XSEDE program,
and by the University of Wuppertal. 
KH wishes to thank the Institute for Theoretical Physics and the Albert Einstein Center for Fundamental Physics at Bern University for their support.
KH and JK wish to thank the Galileo Galilei Institute for Theoretical Physics and INFN for their hospitality and support at the workshop
"New Frontiers in Lattice Gauge Theories".


\end{document}